\documentclass[preprint,aps,nofootinbib]{revtex4-2}
\pdfoutput=1
\usepackage{physics}
\usepackage{graphicx}   
\usepackage{multirow}
\usepackage{hyperref}
\usepackage{color}
\usepackage{amsmath}

\begin{document}

\title{Big bounce and black bounce in quasi-topological gravity}
\author{Yi Ling $^{1,2}$}
\email{lingy@ihep.ac.cn}
\author{Zhangping Yu $^{1,2}$}
\email{yuzp@ihep.ac.cn}
\affiliation{$^1$Institute of High Energy
  Physics, Chinese Academy of Sciences, Beijing 100049, China\\ $^2$
  School of Physics, University of Chinese Academy of Sciences,
  Beijing 100049, China }

\begin{abstract}
In the framework of quasi-topological (QT) gravity, we propose a novel model which is characterized by a bounce of the spacetime such that the singularity in standard general relativity can be avoided in both cosmological and black hole setups. Specifically, in the cosmological background, this model reproduces the modified Friedmann equation proposed in loop quantum cosmology, while in a black hole background, it produces a black bounce metric identical to that of the quantum Oppenheimer-Snyder (qOS) model. This model resolves the singularity presented in the qOS model as well as in QT gravity coupled to linear electromagnetic fields, and provides a unified, manifestly covariant framework for general spacetimes, from which both the modified Friedmann equation and the qOS black hole metric can be derived. Furthermore, it establishes a profound correspondence between the effective dynamics of loop quantum cosmology and the QT gravity theory, suggesting that certain quantum gravitational effects in loop quantum gravity can be captured by adding an infinite tower of higher-curvature corrections to the Einstein-Hilbert action.
\end{abstract}
\maketitle

\newpage

\section{Introduction}
Currently general relativity (GR) as the most successful theory of gravitation faces a key challenge. It predicts the formation of spacetime singularities, such as the Big Bang singularity in cosmology and the singularity at the center of black holes \cite{Penrose:1964wq,Hawking:1973uf}, where physical quantities like spacetime curvature and energy density become infinite, rendering the predictive power of the theory void and implying that its framework breaks down. Consequently, it is widely accepted that a theory of quantum gravity is necessary to resolve these singularities and provide a more fundamental description of spacetime at the Planck scale.

A compelling paradigm that has emerged from various quantum gravity approaches is the replacement of classical singularities with a ``bounce''. In the cosmological context, this implies that the Big Bang is not the beginning of time, but rather a point of phase transition from a preceding contracting universe to our current expanding one--a scenario known as the Big Bounce. The existence of a Big Bounce resolving the Big Bang singularity has obtained diverse support in Loop Quantum Cosmology (LQC) \cite{Bojowald:2001xe,Ashtekar:2006rx,Ashtekar:2006wn,Bojowald:2008zzb,Novello:2008ra,Assanioussi:2018hee}, and a prominent model is the Ashtekar-Pawlowski-Singh (APS) model \cite{Ashtekar:2006rx}. In this model, the classical Friedmann equation is superseded by a modified version that incorporates a critical density at the Planckian order. As the energy density of the universe reaches the critical density, the gravitational force effectively becomes repulsive, causing the universe to ``bounce'' and transit into an expanding phase, thereby averting the singularity.

In the context of black holes, a parallel mechanism is proposed to resolve the singularity within the horizon of black holes. It suggests that infalling matter does not get crushed into a singularity but instead undergoes a quantum bounce at the Planck scale, and eventually re-emerges from a white hole. This process, often referred to as a black-to-white hole transition or a ``black bounce'', leads to the formation of regular black holes. Numerous bouncing models have been investigated in the literature \cite{Rovelli:2014cta,Olmedo:2017lvt,Bianchi:2018mml,DAmbrosio:2018wgv,Ashtekar:2018lag,Ashtekar:2018cay,Simpson:2018tsi,Simpson:2019cer,Gambini:2020nsf,Lobo:2020ffi,Kelly:2020uwj,Kelly:2020lec,Lobo:2020kxn,Ashtekar:2021kfp,Ashtekar:2023cod,Pereira:2023lck,Feng:2023pfq,Feng:2024sdo,Alencar:2024yvh,Pal:2024kng,Rajagopal:2025qyp,Pereira:2025xnw}.

While bouncing models exist for both cosmological and black hole scenarios, several significant challenges remain. First, cosmological and black hole models are typically treated independently due to the distinct symmetry of spacetime inherent to each setup, and a unified framework for general spacetimes has yet to be developed. Second, the majority of these models are formulated within the Hamiltonian framework, and as a consequence, the resultant theory is not manifestly covariant. For instance, in the model proposed in Ref.~\cite{Kelly:2020uwj,Kelly:2020lec}, different choices of the lapse function may lead to physically distinct metrics which are not related by coordinate transformations, explicitly demonstrating the non-covariance of the model~\cite{Kelly:2020uwj}. The quantum Oppenheimer-Snyder (qOS) model~\cite{Lewandowski:2022zce} also lacks manifest covariance; furthermore, its exterior Schwarzschild geometry is not derived from a complete theory but is instead imposed through ad hoc junction conditions. While a covariant Hamiltonian formulation has recently been developed~\cite{Zhang:2024khj,Zhang:2024ney,Yang:2025ufs,Zhang:2025ccx}, this method is currently restricted to spherically symmetric spacetimes. Its generalization to more general spacetimes, as well as to cosmological settings for realizing a big bounce, remains an open question. Finally, even in some models where the black hole interior is non-singular~\cite{Kelly:2020uwj,Kelly:2020lec,Lewandowski:2022zce}, the singularity is not fully resolved, as the maximally extended exterior spacetime still contains another sheet with a singularity at $r=0$ (see Fig.~2(a) of Ref.~\cite{Lewandowski:2022zce}).

Recently, a new strategy for avoiding singularities has been proposed in the framework of Quasi-topological (QT) gravity, which is a class of higher-order gravitational theories that possess particularly desirable properties over the spherically symmetric backgrounds \cite{Myers:2010ru,Kuang:2010jc,Hennigar:2017ego,Bueno:2019ycr,Bueno:2019ltp,Bueno:2022res}. Specifically, the equations of motion remain to be second-order in spherical symmetry and admit a Birkhoff-like theorem, ensuring the uniqueness of the static solution. On one hand, previous research has shown that QT theories containing infinite towers of higher-curvature corrections can resolve the cosmological singularity problem, through a mechanism known as ``geometric inflation'', where the Big Bang is replaced by a period of exponential growth \cite{Arciniega:2018tnn,Arciniega:2018fxj,Edelstein:2020nhg,Arciniega:2020pcy}. It is noteworthy, however, that these models are characterized by an unbounded energy density ($\rho \rightarrow \infty$ as $t \rightarrow -\infty$), {\it which distinguishes them from a typical Big Bounce}. On the other hand, recent studies have demonstrated that when considering a spherically symmetric background spacetime, these same theories can also resolve black hole singularities, yielding regular black holes such as the Hayward model \cite{Bueno:2024dgm,Bueno:2024eig,Bueno:2024zsx,DiFilippo:2024mwm,Frolov:2024hhe,Bueno:2025gjg}. Given that QT gravity can resolve the singularity problem in both black hole and cosmological contexts, it is natural to investigate whether this framework can be applied to derive a cosmological bounce akin to that in LQC, while also producing a ``black bounce'' model for black holes. It motivates us to explore this idea in this work. 

In this paper, we construct a novel bouncing model within the QT gravity, unifying singularity resolution in black hole spacetimes and cosmological scenarios. By proposing a specific QT gravity action, we derive a modified Friedmann equation identical to that in LQC \cite{Ashtekar:2006rx}, and a black bounce metric identical to the quantum-corrected Schwarzschild metric proposed in the qOS model \cite{Lewandowski:2022zce}. The cosmological solution exhibits a critical energy density that triggers a bounce, while the quantum-corrected Schwarzschild spacetime exhibits a minimal radius $r_0$, thereby excluding the central singularity. In both cosmological and black hole scenarios, spacetime exhibits a bounce process. Notably, these results emerge from a purely geometric action principle without ad hoc matter sources. Besides, this model is manifestly covariant, as it is derived from the Lagrangian formalism. It also resolves the singularities in both the maximally extended exterior of the qOS model and QT gravity coupled to linear electromagnetic fields. We further investigate the thermodynamics of the black hole by calculating its Hawking temperature and thermodynamical entropy. We show that the entropy matches the value derived from the quantum dispersion relations \cite{Ling:2009wj}, and satisfies the first law of thermodynamics. Our findings reveal a profound correspondence between QT gravity and LQG at the level of effective dynamics, suggesting that certain quantum gravitational effects in LQG can be realized in QT gravity. 

The paper is organized as follows: In Section II we review the foundation of QT gravity and its application to singularity resolution. In Section III we propose a specific bouncing model within QT gravity, deriving the modified Friedmann equation with big bounce in cosmological setup and the black bounce metric in black hole setup, respectively. The thermodynamical property of the black hole is also analyzed. The conclusion and discussion are given in Section IV.

\section{Quasi-topological Gravities and Singularity Resolution}
In this section we briefly present a review on the framework of QT gravity. Consider a general theory constructed from arbitrary contractions of the Riemann tensor and the metric, with the Lagrangian density $\mathcal{L}(g^{ab},R_{cdef})$,
\begin{equation}
  S = \frac{1}{16\pi G}\int \dd[D]{x} \sqrt{-g} \mathcal{L}(g^{ab},R_{cdef}).
\end{equation}
The equations of motion of this theory in the absence of matter are given by \cite{Padmanabhan:2011ex}
\begin{equation}
  P_{a}^{cde}R_{bcde}-\frac{1}{2}g_{ab}\mathcal{L}-2\nabla^{c} \nabla^{d}P_{acdb}=0 ,
\end{equation}
where $P^{abcd} \equiv \partial{\mathcal{L}}/ \partial{R}_{abcd}$. Note that the last term $\nabla^{c} \nabla^{d}P_{acdb}$ generally introduces higher-order derivatives of the metric. According to Ostrogradsky's theorem \cite{Ostrogradsky:1850fid,Woodard:2015zca}, such terms typically lead to pathological instabilities, resulting in a Hamiltonian that is linear in the canonical momenta, and consequently an energy that is unbounded from below. This instability can be avoided by imposing the condition $\nabla^{d}P_{acdb}=0$, which prevents the introduction of higher-order derivatives and defines a class of theories known as Lovelock theories \cite{Lovelock:1970zsf,Lovelock:1971yv,Padmanabhan:2013xyr}. However, in a $D$-dimensional spacetime, it is not possible to define a Lovelock theory containing curvature terms of an arbitrarily high order. In fact, the Lovelock theories can only be defined up to orders $n \le \lfloor D/2 \rfloor$.

Quasi-topological (QT) gravity is a generalization of Lovelock theory that allows for curvature terms of arbitrarily high order \cite{Myers:2010ru,Hennigar:2017ego,Bueno:2019ycr,Bueno:2019ltp,Bueno:2022res}. It achieves this by relaxing the stringent conditions of Lovelock theory, requiring that $\nabla^{d}P_{acdb}$ vanishes only over the spherically symmetric spacetime:
\begin{equation}
  \nabla^{d}P_{acdb} \left|_{\text{SS}}\right.=0,
\end{equation}
where $\left|_{\text{SS}}\right.$ denotes that this expression is evaluated on a spherically symmetric metric
\begin{equation}
  \label{eq:SSmetric}
  \mathrm{d}s_{\mathrm{SS}}^{2}=-N(t,r)^{2}f(t,r)\mathrm{d}t^{2}+\frac{\mathrm{ d}r^{2}}{f(t,r)}+r^{2}\mathrm{d}\Omega_{(D-2)}^{2} .
\end{equation}

In spacetimes with $D \ge 5$ dimensions, QT gravity theories can be constructed from a linear combination of $n$-th order curvature invariants $\mathcal{Z}_{(n)}$:
\begin{equation}
  \label{eq:action}
  S = \frac{1}{16\pi G}\int \dd[D]{x} \sqrt{-g} \left(R + \sum_{n=2}^\infty a_n \mathcal{Z}_{(n)} \right),
\end{equation}
where $a_n$ are coupling constants with dimensions of $[\text{length}]^{(2n-2)}$. The explicit expressions for $Z_{(2)} \sim Z_{(5)}$ are provided in Refs.~\cite{Bueno:2024eig,Bueno:2024zsx,Bueno:2025gjg}, and higher-order $\mathcal{Z}_{(n)}$ terms can be built via the recursion relation \cite{Bueno:2019ycr}
\begin{equation}
  \mathcal{Z}_{(n+5)}=\frac{3(n+3)\mathcal{Z}_{(1)}\mathcal{Z}_{(n+4)}}{D(D-1)( n+1)}-\frac{3(n+4)\mathcal{Z}_{(2)}\mathcal{Z}_{(n+3)}}{D(D-1)n}+\frac{(n+3)(n+4) \mathcal{Z}_{(3)}\mathcal{Z}_{(n+2)}}{D(D-1)n(n+1)} .
\end{equation}

The equations of motion for action (\ref{eq:action}) under the metric ansatz (\ref{eq:SSmetric}) can be obtained by taking the variation of the reduced action $S_{2d} = S|_{\text{SS}}$ with respect to $N(t,r)$ and $f(t,r)$. It is found that this reduced action $S_{2d}$ corresponds to a two-dimensional Horndeski tensor theory \cite{Horndeski:1974wa}, whose equations of motion are remarkably simple. Including a stress-energy tensor $T_{ab}$ in the action Eq.~\eqref{eq:action}, one finally obtains the equations of motion for $N(t,r)$ and $f(t,r)$ as follows:
\begin{equation}
  \begin{aligned}\label{eq:SSeom}
    \begin{aligned}
\partial_r\left[r^{D-1}h(\psi)\right] & =\frac{16\pi G}{(D-2)N^2f}r^{D-2}T_{tt}, \\
\partial_{t}f & =-\frac{16\pi G}{(D-2)h^{\prime}(\psi)}rfT_{tr}, \\
\partial_r N & =\frac{8\pi G}{(D-2)h^{\prime}(\psi)}rN\left(T_{rr}+\frac{1}{N^2f^2}T_{tt}\right).
\end{aligned}
  \end{aligned}
\end{equation}

where
\begin{equation}
  \label{eq:handpsi}
  h(\psi) \equiv \sum_{i=1}^\infty a_i \psi^i,\quad a_1 = 1,\quad \psi \equiv \frac{1-f(r)}{r^2}.
\end{equation}
In the vacuum case where $T_{ab}=0$, the solutions to Eqs.~(\ref{eq:SSeom}) are
\begin{equation}
  N(t,r) = 1, \quad h(\psi) = \frac{2 M}{r^{D-1}}, \quad f(t,r) =  1 - r^2 h^{-1}\qty(\frac{2M}{r^{D-1}}) ,
\end{equation}
and $M$ is an integration constant which is related to the ADM mass of the black hole $m$ through
\begin{equation}
  {M}\equiv\frac{8\pi G m}{(D-2)\Omega_{(D-2)}},
\end{equation}
where $\Omega_{(D-2)}=2\pi^{(D-1)/2}/\Gamma\qty[(D-1)/2]$ is the volume of the $(D-2)$ sphere. The characteristic function $h(\psi)$ encodes a great deal of information about QT gravity.

It was proposed in Ref.~\cite{Bueno:2024dgm} that if the Lagrangian contains an infinite tower of higher curvature terms, and the coefficients $a_n$ satisfy
\begin{equation}
  \label{eq:RBHcondition}
  a_{n}\geq 0\ \forall n ,\quad\lim_{n\to\infty}(a_{n})^{\frac{1}{n} }=C>0 ,
\end{equation}
then the solutions to the equations of motion describe regular black holes. A simple example is obtained by setting $a_n = \alpha^{n-1}$, which leads to a generalized Hayward black hole with $f(r) = 1-2Mr^{2}/\qty(r^{D-1}+ 2\alpha M)$. It is worth noting that condition (\ref{eq:RBHcondition}) is sufficient but not necessary. 

When coupled to Maxwell electromagnetic field, the stress-energy tensor is given by
\begin{equation}
  T_{tt} = -N^2 f^2 T_{rr} = -\frac{(D-2)(D-3)}{16\pi G} \frac{Q^2}{r^{2D-4}}N^2 f, \quad T_{tr} = 0,
\end{equation}
where $Q$ denotes the total electric charge of the system. We still have $N(t,r) = 1$, and the resulting charged metric function takes the form
\begin{equation}
  f(r) = 1 - {r^2} h^{-1}\qty(\frac{2M}{r^{D-1}}-\frac{Q^2}{r^{2D-4}}),
\end{equation}
which corresponds to replacing $M$ with the Misner-Sharp mass $M-Q^2 / 2r^{D-3}$ in the vacuum metric function. Note that both the Misner-Sharp mass and the stress–energy tensor diverge as $r \to 0$. Consequently, the charged metric and the electromagnetic energy density exhibit singularities at finite radius. This highlights a generic difficulty encountered in constructing regular charged black holes in QT gravity: even if the QT gravity theory admits regular vacuum solutions, coupling to linear Maxwell electrodynamics typically destroys regularity. To avoid this, one is usually forced to invoke nonlinear electrodynamics, such as the Born–Infeld model \cite{Hennigar:2025yqm}. In next section, we will present a model which may offer an alternative resolution without modifying the linearity of electrodynamics. 

QT gravity can also resolve the cosmological singularity problem. Consider a Friedmann-Lemaître-Robertson-Walker (FLRW) ansatz
\begin{equation}
  \mathrm{d}s^{2}=-\mathrm{d}\tau^{2}+a(\tau)^{2}\left[\frac{\mathrm{d}r^{2} }{1-k r^{2}}+r^{2}\mathrm{d}\Omega_{(D-2)}^{2}\right] ,
\end{equation}
where $k=0,1,-1$ corresponds to flat, closed, and open universes, respectively. The equations of motion for the action in Eq.~(\ref{eq:action}) under the FLRW ansatz can be derived through an analogous procedure. Under this ansatz, the action again reduces to a two-dimensional Horndeski theory, yielding the corresponding equations of motion:
\begin{equation}
  \label{eq:FLRWeom}
  h(\Phi)=\varrho ,\quad\Phi\equiv\frac{k+\dot{a}(\tau)^{2}}{a( \tau)^{2}} ,
\end{equation}
and
\begin{equation}
  \varrho\equiv\frac{16\pi G \rho}{( D-2 )( D-1 )} .
\end{equation}
On the other hand, from the conservation equation for the stress-energy tensor of the fluid, one obtains
\begin{equation}
  \dot{\rho}+(D-1)(\rho+p)\frac{\dot{a}(\tau)}{a(\tau)}=0.
\end{equation}
Upon specifying an equation of state for the matter, the scale factor $a(t)$ can be obtained by solving above two equations.

The equations of motion (\ref{eq:FLRWeom}) were first derived and solved in Ref.~\cite{Arciniega:2018tnn} \footnote{As standard quasi-topological gravity is trivial in four dimensions, Ref.~\cite{Arciniega:2018tnn} is actually concerned with the generalized quasi-topological gravity \cite{Bueno:2019ycr}, which yields identical equations of motion Eq.~(\ref{eq:FLRWeom}) under the FLRW ansatz.}, where two (generalized) quasi-topological theories with different coupling constants $a_n$ were analyzed. Both models contain infinite towers of higher-curvature corrections, and the solutions of Eq.~(\ref{eq:FLRWeom}) indicate that the scale factor $a(t)$ scales as $\exp(\alpha\, t^\beta)$, where $\alpha$ and $\beta$ are constants. Therefore, for these models, $a \rightarrow 0$ is only reached asymptotically as $t \rightarrow -\infty$. The resulting spacetime is thus geodesically complete and free of singularities, and there is no notion of a temporal origin at which we can set $t=0$. This exponential behavior of $a(t)$ is a general feature of various QT gravity models with different coupling coefficients $a_n$, and this scenario has been termed ``geometric inflation''. However, not all QT gravity theories incorporating an infinite number of terms lead to geometric inflation. In fact, the model we present in this work serves as a counterexample; instead of geometric inflation, it yields a Big Bounce.

A final remark on the dimensionality of spacetime is in order. The QT gravity theory underpinning our work, which involves an infinite number of curvature terms, is well-defined only in $D \ge 5$ dimensions, as it reduces to standard Einstein gravity in $D = 4$. Accordingly, our analysis in next section is valid for any spacetime dimension $D \ge 5$. Nevertheless, to facilitate a direct comparison with results from loop quantum gravity, we will extend our discussion to the $D=4$ case. This extension is well-justified because all subsequent calculations rely only on the equations of motion, Eqs. (\ref{eq:SSeom}) and (\ref{eq:FLRWeom}), irrespective of the specific underlying Lagrangian. Although a quasi-topological theory with infinite terms cannot be defined in $D=4$, a generalized quasi-topological theory exists in four dimensions whose FLRW equation of motion is precisely Eq.~(\ref{eq:FLRWeom}), even though Eq.~(\ref{eq:SSeom}) no longer holds \cite{Bueno:2019ycr}. More compellingly, a four-dimensional scalar-tensor theory has been proposed in \cite{Fernandes:2025fnz} where both Eqs.~(\ref{eq:SSeom}) and (\ref{eq:FLRWeom}) are satisfied simultaneously, implying the identical equations of motion in QT gravity with $D \ge 5$ could be derived in four dimensions from other alternative theories. This provides a solid foundation for applying our analysis of the big bounce and black bounce to the four-dimensional case.

\section{Bounce in quasi-topological gravities}
In this section, we propose a model characterized by a bounce in the framework of QT gravity. We begin by defining the characteristic function $h(x)$ as follows:
\begin{align}
  \label{eq:hxminus}
  h(x) & = \frac{\alpha - \sqrt{\alpha^2 - 4 \alpha l^2 x}}{2 l^2} \nonumber                                 \\
       & =\sum_{n=0}^\infty\left.\frac{1}{n!}\frac{\dd^{n} h(x)}{\dd x^n}\right|_{x=0}x^n \nonumber          \\
       & = x + \frac{l^2 }{\alpha} x^2 + \frac{2 l^4 }{\alpha^2} x^3 + \frac{5 l^6 }{\alpha^3} x^4 + \ldots,
\end{align}
where $\alpha=16\pi G / \qty[( D-2 )( D-1 )]$ and $l=\sqrt{\qty(32 \pi^2 \gamma^3 G^2 \hbar) / \sqrt{3}}$ with the Barbero-Immirzi parameter $\gamma$. That is to say, we consider a QT gravity with an action
\begin{equation}
  \label{eq:Sminus}
  S = \frac{1}{16\pi G}\int \dd[D]{x} \sqrt{-g} \left(R + \frac{l^2 }{\alpha} \mathcal{Z}_{(2)} + \frac{2 l^4 }{\alpha^2} \mathcal{Z}_{(3)} + \frac{5 l^6 }{\alpha^3} \mathcal{Z}_{(4)} + \ldots\right).
\end{equation}
Firstly, we consider the cosmological solutions in this QT gravity. According to Eq.~(\ref{eq:FLRWeom}), the equation of motion for the FLRW metric corresponding to this action is
\begin{equation}
  h(\Phi) = \frac{\alpha - \sqrt{\alpha^2 - 4 \alpha l^2 \Phi}}{2 l^2} = \varrho = \alpha \rho,
\end{equation}
and we get
\begin{equation}
  \label{eq:FLRWsolution}
  \Phi = \frac{k + \dot{a}(\tau)^2}{a(\tau)^2} = \alpha\rho\left(1-l^2 \rho\right) = \frac{16\pi G}{( D-2 )( D-1 )} \rho (1-\frac{\rho}{\rho_{c}}) ,
\end{equation}
where $\rho_{c} = 1/l^2 = \sqrt{3}/(32 \pi^{2}\gamma^{3}G^2 \hbar)$ is the critical density. In the case of $k=0$, this relation reproduces the modified Friedmann equation of higher-dimensional LQC \cite{Zhang:2015bxa}. Specifically, in the limit $D=4$, it recovers the standard result established in the APS model \cite{Ashtekar:2006rx}
\begin{equation}
  \label{eq:modifiedFriedmann}
  H^2 \equiv \frac{\dot{a}}{a}  = \frac{8\pi G}{3} \rho \left(1-\frac{\rho}{\rho_{c}}\right) ,
\end{equation}
Furthermore, for $k = \pm 1$, Eq.~(\ref{eq:FLRWsolution}) yields the generalized expression
\begin{equation}
  H^2 = \frac{16\pi G}{( D-2 )( D-1 )} \rho \left(1-\frac{\rho}{\rho_{c}}\right) - \frac{k}{a(\tau)^2},\label{eq:modifiedFriedmannOpenAndClose}
\end{equation}
describing closed and open universes, respectively.

\begin{figure}[htbp]
    \centering
    \includegraphics{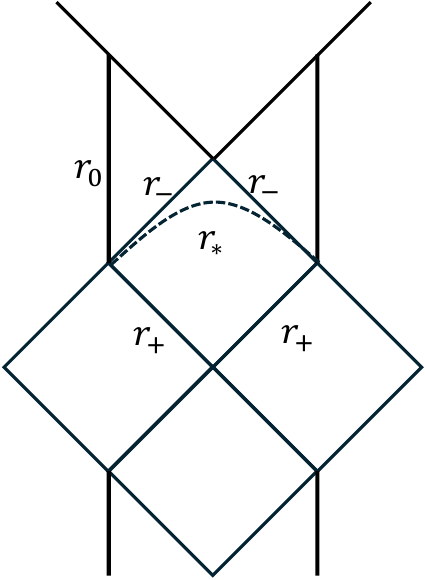} 
    \caption{Penrose diagram of the metric in Eq.~(\ref{eq:qOSmetric}). The classical Schwarzschild singularity is replaced by a region with the minimal radius $r_0$, and $r_*$ denotes the boundary between two regions described by $S_+$ and $S_-$.}
    \label{fig:PenroseDiagram}
\end{figure}

Next we consider the possible black hole solutions in this QT gravity with the same characteristic function $h(x)$. Under the spherically symmetric metric ansatz (\ref{eq:SSmetric}) and following the equations of motion (\ref{eq:SSeom}), we obtain
\begin{align}
  \label{eq:SSsolution}
  f(r) & = 1 - \frac{2 M}{r^{D-3}}+\beta\frac{M^2}{r^{2D-4}},
\end{align}
where $\beta = 4 l^2 / \alpha = 16 \sqrt{3} \pi \gamma^3 G \hbar$. In the case of $D=4$, this leads to the metric
\begin{equation}
  \label{eq:qOSmetric}
  \dd[2]{s} = -\qty(1 - \frac{2G m}{r} + \frac{\beta G^2 m^2}{r^4}) \dd[2]{t} + \qty(1 - \frac{2G m}{r} + \frac{\beta G^2 m^2}{r^4})^{-1}\dd[2]{r} + r^2 \dd[2]{\Omega} .
\end{equation}
The Penrose diagram for this metric is shown in Fig.~\ref{fig:PenroseDiagram}. This metric is consistent with the quantum-corrected Schwarzschild metric derived from the quantum Oppenheimer-Snyder (qOS) model in Refs.~\cite{Lewandowski:2022zce,Shi:2024vki}. In that work, the interior of a spherical star is described by an FLRW metric, with the scale factor satisfying the modified Friedmann equation Eq.~(\ref{eq:modifiedFriedmann}). The exterior metric, given by Eq.~(\ref{eq:qOSmetric}), is then determined by applying junction conditions. To surprise, here we have presented a derivation of the metric from the Lagrangian formalism, and unified the treatment of spherically symmetric black holes and cosmology: from a single Lagrangian, we obtain both the modified Friedmann equation and the quantum-corrected Schwarzschild metric by applying the cosmological and spherically symmetric ansatz, respectively.

Since this black hole metric is derived directly from the equations of motion within the Lagrangian formalism, one can now investigate its thermodynamical behavior following the Wald formalism. The thermodynamics of QT gravity for an arbitrary characteristic function $h(x)$ has been studied in Refs.~\cite{Bueno:2019ycr,Bueno:2022res}, where the Wald entropy reads:
\begin{equation}
  \label{eq:entropy}
  S = -\frac{(D - 2)\Omega_{D-2}}{8G} \int \frac{h'(\psi_{+})}{\psi_{+}^{D/2}} \dd \psi_{+},
\end{equation}
where $\psi_+ = \psi(r_+)$ is the value of $\psi$ at the outer horizon $r_+$. Substituting Eq.~(\ref{eq:hxminus}) and Eq.~(\ref{eq:SSsolution}) into Eq.~(\ref{eq:entropy}), and setting $D=4$, we obtain:
\begin{equation}
  \label{eq:entropyresult}
  S = \frac{A}{4 G}\sqrt{1-\frac{4\pi \beta}{A}} + \frac{\pi\beta}{G} \tanh^{-1}\qty(\sqrt{1-\frac{4\pi \beta}{A}}).
\end{equation}
Here, $A = 4\pi r_+^2$ is the area of the outer horizon. Interestingly, this entropy agrees with the result in Ref.~\cite{Ling:2009wj}. In that work, a modified dispersion relation (MDR) of the form $\sin(\eta l_pE) / \qty(\eta l_p) =\sqrt{p^2+m_0^2}$ was proposed at the phenomenological level of quantum gravity. This led to a corrected entropy-area relation Eq.~(\ref{eq:entropyresult}), which was then used to derive the modified Friedmann equation Eq.~(\ref{eq:modifiedFriedmann}).

Furthermore, the mass and Hawking temperature can be expressed as
\begin{align}
  M &=\frac{r_+^3-r_+^2\sqrt{r_+^2-\beta}}{\beta},\\
  T &= \frac{1}{4\pi}f'(r_+) = \frac{1}{\pi r_+} + \frac{3\qty(-r_+ + \sqrt{r_+^2-\beta})}{2\pi\beta}.
\end{align}
One can straightforwardly verify that this black hole solution, obtained within QT gravity, satisfies the first law of thermodynamics $\dd M=T \dd S$, as pointed out in Ref.~\cite{Bueno:2024dgm}.

So far, we have studied the thermodynamic properties of the metric in (\ref{eq:qOSmetric}). Another important issue to address is the singularity, as this metric appears to be singular at $r=0$. However, we remark that $r=0$ does not fall into the domain of definition and thus is not a point in the spacetime manifold. In the qOS model in Ref.~\cite{Lewandowski:2022zce}, for instance, the radial coordinate is restricted to the region $r \ge r_b$, where $r_b = \qty(8\sqrt{3}\pi \gamma^3 G^2 \hbar m)^{1/3}$ is the stellar radius \footnote{In the qOS model, despite the requirement that $r$ must be greater than the stellar radius $r_b$ , it still exhibits a singularity at $r=0$ in the maximally extended spacetime for the exterior region (see Fig.~2(a) of Ref.~\cite{Lewandowski:2022zce}). This problem is absent in our model. Here, the condition $r>r_0$ is a constraint imposed by the field equations themselves. Consequently, the region $r<r_0$ is excised from the spacetime manifold, rendering it regular everywhere.}. In our model, the allowed range for $r$ is determined by the characteristic function $h(x)$ and the equations of motion. The series expansion Eq.~(\ref{eq:hxminus}) of $h(x)$ converges for $x \in \qty(-\alpha / 4 l^2, \alpha / 4 l^2)$, which implies
\begin{equation}
  -\frac{\alpha}{4l^2} \le \psi = \frac{1-f(r)}{r^2} \le \frac{\alpha}{4l^2}\quad\Rightarrow\quad r \ge r_0 = \qty[\qty(\sqrt{2}-1)\beta M]^{\frac{1}{D-1}}.
\end{equation}
When $r < r_0$, the characteristic function diverges, and one fails to derive the equation of motion Eq.~(\ref{eq:SSeom}). This implies that the region $r < r_0$ is not part of the spacetime, and the metric is therefore regular everywhere. This result is expected, as noted in Refs.~\cite{Kelly:2020uwj,Kelly:2020lec}, since in spherical symmetry there are no gravitational waves, and therefore a central potential must be generated by some distribution of matter. Furthermore, given that the energy density of matter field is bounded by the Planck scale $\rho \lesssim \rho_\text{Pl}$, where $\rho_\text{Pl}$ is the Planck energy density, a source of mass $M$ requires a matter distribution extending out to $r \sim \qty(M/\rho_\text{Pl})^{1/3} \sim r_0$, forming a ``Planck star''\cite{Rovelli:2014cta}. Consequently, a complete description of the spacetime solution must incorporate matter fields, for which the qOS model serves as a concrete example.

An interesting property of the metric (\ref{eq:qOSmetric}) is that the maximal value of the Kretschmann scalar $K^2\left|_{max}\right.\sim M^2 / r_0^{2(D-1)}\sim\rho_\text{Pl}$, is independent of the mass $M$. Furthermore, the existence of a minimum value for $r$ distinguishes this metric from other regular metrics, such as the Hayward metric. It is, in fact, a bouncing solution: a particle in radial motion bounces upon reaching $r=r_0$ , instead of passing through the origin $r=0$.

The analysis of singularities presented herein can be extended to the electromagnetic vacuum. When coupled to linear Maxwell electrodynamics, the charged metric function is obtained by substituting $M$ with $M-Q^2/2r^{D-3}$ in Eq.~\eqref{eq:SSsolution}, which yields
\begin{equation}
f(r) = 1-\frac{2M}{r^{D-3}} + \frac{Q^2}{r^{D-2}} + \frac{\beta}{r^{2D-4}}\qty(M - \frac{Q^2}{2r^{D-3}})^2.
\end{equation}
As previously discussed, the domain of convergence imposes a strict constraint $r \ge r_0$, where $r_0$ satisfies $\psi(r_0)= -\alpha / 4 l^2$. Thus, both the metric and the stress-energy tensor remain regular throughout the entire spacetime. Moreover, in this scenario, the maximum values of the Kretschmann scalar are of the Planck scale and, crucially, are independent of both the mass $M$ and the charge $Q$.\footnote{More precisely, $\eval{K^2}_{\text{max}} = \eval{K^2}_{r=r_0}\sim\rho_\text{Pl} + \mathcal{O}\qty(M^{-\alpha}, Q^{-\beta})$, where $\alpha$ and $\beta$ are positive constants.} This universality of the curvature bound is absent in models based on nonlinear electrodynamics

Following the analysis of the spherically symmetric case, we now turn to examine the constraints imposed by the domain of convergence in the cosmological scenario. It is clear from Eq.~(\ref{eq:modifiedFriedmann}) that $\Phi = H^2 \le \frac{\alpha}{4 l^2}$, which ensures that the series in Eq.~(\ref{eq:hxminus}) is always convergent.

Although the properties of this QT gravity model have been extensively studied, it is important to pointed out that the spacetime, derived from the characteristic function (\ref{eq:hxminus}) and the action (\ref{eq:Sminus}), does not cover the entire region $r \in [r_0,\infty)$, but only a portion of it. The reason is that the characteristic function is restricted by the inequality  
\begin{equation}
  h(x) = \frac{\alpha - \sqrt{\alpha^2 - 4 \alpha l^2 x}}{2 l^2} \le \frac{\alpha}{2l^2}.
\end{equation}
It follows that
\begin{gather}
  h(\psi) = \frac{2 M}{r^{D-1}} \le \frac{\alpha}{2l^2}\quad\Rightarrow\quad r\ge r_* = \qty(\frac{4 l^2 M}{\alpha})^{\frac{1}{D-1}},\\
  h(H^2) = \alpha \rho \le \frac{\alpha}{2l^2} \quad\Rightarrow\quad \rho \le \frac{\rho_c}{2}.
\end{gather}
Therefore, the spacetime derived from $S$ describes only the spacetime region $r \in [r_*,\infty)$ or $\rho \in [0,\rho_c/2]$. Henceforth, we shall denote the characteristic function (\ref{eq:hxminus}) and the action (\ref{eq:Sminus}) as $h_-(x)$ and $S_-$, respectively. To cover the region $r \in [r_0, r_*]$ in the black hole setup and $\rho \in [\frac{\rho_c}{2}, \rho_c]$ in the cosmological setup, we introduce the characteristic function $h_+(x)$ and the action  $S_+$ as follows.
\begin{align}
  h_{\pm} &= \frac{\alpha \pm \sqrt{\alpha^2 - 4 \alpha l^2 x}}{2 l^2} = \sum_{n=0}^{\infty}a_n^{\pm} \; x^n,\\
  S_{\pm} &= \frac{1}{16\pi G}\int \dd[D]{x} \sqrt{-g} \left(\sum_{n=0}^\infty a_n^{\pm} \mathcal{Z}_{(n)}\right).
\end{align}
It can be shown that the pair of characteristic functions and actions yield the same metric. Specifically, the field equations and corresponding solutions under the spherically symmetric and FLRW ansatz are as follows:
\begin{gather}
  h_{\pm}(\psi) = \frac{2 M}{r^{D-1}} \quad \Rightarrow\quad f(r) = 1 - \frac{2 M}{r^{D-3}}+\beta\frac{M^2}{r^{2D-4}},\\
  h_{\pm}(H^2) = \alpha \rho \quad\Rightarrow \quad H^2 = \alpha \rho \left(1-\frac{\rho}{\rho_{c}}\right).
\end{gather}
Although the field equations derived from these two actions yield identical solutions, the actions themselves actually describe different regions. Correspondingly, since $h_+(x)\ge \alpha /2l^2$, the spacetime derived from $S_+$ is restricted to the region $r \in [r_0,r_*]$ or $\rho \in [\rho_c/2,\rho_c]$. The differences between these two actions are summarized in Table \ref{tab:actions}.

\begin{table}[htbp]
\centering
\caption{Comparison of the two actions $S_+$ and $S_-$}
\label{tab:actions}
\renewcommand{\arraystretch}{1.15}
\setlength{\tabcolsep}{14pt} 
\begin{tabular}{|c|c|c|}
\hline
 & $S_+$ & $S_-$ \\
\hline
$h(x)$ & $h_+(x)$  & $h_-(x)$ \\
\hline
\begin{tabular}{@{}c@{}} The equations of \\ motion \end{tabular} & $\begin{gathered}
  h_+(\psi) = \frac{2 M}{r^{D-1}} \\
  h_+(H^2) = \alpha \rho
\end{gathered}$ & $\begin{gathered}
  h_-(\psi) = \frac{2 M}{r^{D-1}} \\
  h_-(H^2) = \alpha \rho
\end{gathered}$\\
\hline
Solutions & \multicolumn{2}{c|}{
  $\begin{gathered}
  f(r) = 1 - \frac{2 M}{r^{D-3}}+\beta\frac{M^2}{r^{2D-4}} \\
  H^2 = \alpha \rho \left(1-\frac{\rho}{\rho_{c}}\right)
  \end{gathered}$
} \\
\hline
\begin{tabular}{@{}c@{}} The domain of \\ convergence of $h(x)$ \end{tabular} 
        & \multicolumn{2}{c|}{$|x| <\frac{\alpha}{4l^2} \quad \Rightarrow\quad r\ge r_0$} \\
\hline
The range of $h(x)$ & $h_+(x)\ge \alpha /2l^2$ & $h_-(x)\le \alpha /2l^2$\\
\hline
Cover region & $\begin{gathered}
  r \in [r_0, r_*] \\
  \rho \in [\frac{\rho_c}{2}, \rho_c]
  \end{gathered}$ & $\begin{gathered}
  r \in [r_*, +\infty) \\
  \rho \in [0, \frac{\rho_c}{2}]
  \end{gathered}$\\
\hline
\end{tabular}
\end{table}

In summary, a complete description of the spacetime requires both actions, with the total action given by
\begin{equation}
  S = S_+ + S_-.
\end{equation}
Neither action alone is sufficient. The action $S_+$ describes the quantum-dominated region where the bounce occurs, characterized by a Kretschmann scalar $K^2 \sim \rho_\text{Pl}$ or an energy density $\rho \sim \rho_\text{Pl}$. Conversely, $S_-$ describes the region dominated by classical effects, which does not contain the bounce region. Both actions are essential for deriving the bounce solutions. This dual-action structure is analogous to the approach in Ref.~\cite{Ling:2009wj}, where two dispersion relations between $\delta E$ and $\delta p$ are both required to obtain the modified Friedmann equation.

Finally, we may point out that the necessity for two actions stems from the mathematical structure of the characteristic functions $h_{\pm}(x)$. Both $h_{\pm}(x)$ are solutions to the single quadratic equation $x=h-l^2 h^2/\alpha\equiv g(h)$. In another word, the characteristic function $h(x)$ as the inverse function of the single-valued function $g$ has two branches, which is a double-valued function. Thus one needs to consider both of them. In the absence of quantum corrections, then this equation is simply $x=h$, which gives rise to the results in standard GR. Note that for a given metric function $f(r)=1-r^2 g(2M/r^{D-1})$ with an arbitrary function $g$, the metric can be derived from the single action in Eq.~\eqref{eq:action} only if $g$ is monotonic.  When $g$ is non-monotonic, the action must consist of multiple branches, as is the case here. Interestingly, \cite{MyUpcomingPaper} highlights a correspondence between the covariant framework of effective quantum gravity and QT gravity. While the latter requires two branches to describe the entire spacetime, the covariant framework employs only a single effective Hamiltonian \cite{Zhang:2025ccx}
\begin{equation}
  \label{eq:Hamiltonian}
  H_\text{\rm eff} = -2 E^2 \qty(\partial_{s_1} M_{\rm eff} + \frac{s_3}{2} \partial_{s_2} M_{\rm eff} + \frac{s_5}{s_4} \partial_{s_4} M_{\rm eff}),
\end{equation}
with mass function
\begin{equation}
  M_{\rm eff}(s_1, s_2, s_4) = \frac{\sqrt{s_1}^3}{2\zeta^2}\sin^2\qty(\frac{\zeta s_2}{\sqrt{s_1}} \pm \frac{2 \zeta \Xi }{s_1}) \mp \frac{s_1 \sqrt{\qty(s_4)^2 - 4} \sin \qty(\frac{2\zeta s_2}{\sqrt{s_1}} \pm \frac{4 \zeta \Xi }{s_1})}{4\zeta},
\end{equation}
where $s_i$ ($i=1, \dots, 5$) are functions of the canonical variables $E^I$ and $K_I$ ($I=1, 2$), $\Xi = \sqrt{s_1 s_4^2 - 4 s_1} / 4$, and $\zeta$ is a constant characterizing the quantum effects. In this framework, the modified Friedmann equations are expressed in terms of $g$ rather than $h$, specifically $H^2 = g(\alpha \rho)$. This suggests that the Hamiltonian formalism may offer a more unified and advantageous description of the global spacetime structure.

From a phenomenological perspective, this quadratic equation $x=h-l^2 h^2/\alpha$ has been widely considered to generalize the uncertainty principle (GUP) and modify dispersion relations (MDR) at the phenomenological level such that the divergent behavior of the Hawking temperature could be avoided at the final stage of the black hole evaporation \cite{Adler:2001vs,Ling:2005bq,Han:2008sy}. The big bounce solutions are also obtained with the help of MDR in the framework of rainbow gravity \cite{Ling:2008sy}.

\section{Conclusion and Discussion}
In this paper, we have constructed a novel model within the framework of QT gravity that successfully unifies the resolution of the singularity in both black hole and cosmological settings. By proposing specific characteristic functions $h_\pm(x)$ inspired by LQC, we have demonstrated that a purely geometric action can yield the effective dynamics of both a bouncing universe and a bouncing black hole. For the FLRW metric, our model reproduces the well-known modified Friedmann equation from the APS model, which replaces the Big Bang singularity with a Big Bounce. For a spherically symmetric spacetime, the same action leads to a black bounce metric, identical to the quantum-corrected Schwarzschild solution derived from the qOS model. This solution is regular everywhere, with the central singularity being avoided by a bounce at a minimum radius $r_0$. We have further shown that the black hole entropy matches the result derived from quantum dispersion relations, and the first law of thermodynamics is satisfied. Although the formalism of QT gravity is rigorously defined for dimensions $D\ge 5$, we have argued for the validity of $D=4$ results since the same calculation can also be applied to the scalar-tensor models where the fundamental equations of motion Eq.~(\ref{eq:SSeom}) and Eq.~(\ref{eq:FLRWeom}) still hold \cite{Fernandes:2025fnz}. Furthermore, while our investigation has been confined to the spherically symmetric and FLRW ansatz, the action of QT gravity is universal and can be applied to spacetimes without specific symmetry assumptions. Our work reveals a profound correspondence between the effective dynamics of LQC and a specific class of infinite-order QT gravity theories, suggesting that certain quantum gravitational effects in LQG can be captured by adding higher-curvature corrections. 

The modified Friedmann equation for $k=\pm1$ derived here, namely Eq.~\eqref{eq:modifiedFriedmannOpenAndClose}, warrants particular attention. In LQC, the effective dynamics for the flat universe ($k=0$), described by Eq.~\eqref{eq:modifiedFriedmann}, are well-established; however, no consensus exists for the curved cases ($k=\pm1$), where different quantization schemes yield disparate dynamics. It is noteworthy that while our $k=0$ result recovers the standard LQC equation, our $k=\pm1$ formulation deviates from existing LQC proposals. Despite this discrepancy, our approach offers a crucial advantage: within the context of the quantum Oppenheimer-Snyder model, employing our modified Friedmann equation for the interior geometry yields a unique exterior static metric, Eq.~\eqref{eq:qOSmetric}, independent of the spatial curvature $k$. In contrast, utilizing alternative LQC prescriptions for $k=\pm1$ results in exterior metrics that fail to align with the flat case \cite{Lewandowski:2022zce}. Furthermore, given the correspondence between QT gravity and the covariant framework \cite{MyUpcomingPaper}, our equations for $k=\pm1$ are consistent with the formulation derived therein.

It is instructive to compare our model with the QT gravity characterized by coefficients $a_n = \alpha^{n-1}$. As mentioned, this theory yields a generalized Hayward black hole in the spherically symmetric case \cite{Bueno:2024dgm}, and a geometric inflation solution in the $k=0$ cosmological case \cite{Arciniega:2018tnn}, neither of which incorporates a bouncing process. However, as shown in Ref.~\cite{Bueno:2025gjg}, the Oppenheimer-Snyder collapse within that same theory --- which utilizes a $k=1$ FLRW metric in the black hole interior --- does undergo an infinite number of bounces. The crucial distinction lies in the role of spatial curvature $k$. For the choice $a_n = \alpha^{n-1}$, QT gravity yields a geometric inflation metric in a flat ($k=0$) universe but a bouncing metric in a closed ($k=1$) one. In contrast, our proposed model exhibits bouncing behavior universally: for the spherically symmetric black hole scenario and for all cosmological cases ($k=0,\pm1$). 

A notable feature of our model is the necessity of employing two branches of the action, $S_+$ and $S_-$, to describe the entire spacetime. The $S_+$ branch governs the high-curvature regime where quantum effects dominate and the bounce occurs, while the $S_-$ branch describes the low-curvature, quasi-classical region. This is analogous to Ref.~\cite{Ling:2009wj}, which also requires two distinct relations between $\delta E$ and $\delta p$. Despite the necessity for a dual-branched action, as noted in the previous section, a single Hamiltonian Eq.~\eqref{eq:Hamiltonian} suffices to describe the entire spacetime evolution. In light of this, and considering that two dispersion relations are derived from the unique condition $\sin(\eta l_pE) / \qty(\eta l_p) =p$, it is plausible that the total action $S=S_+ + S_-$ also admits a unified representation. The feasibility of this unification hinges on whether the series $\sum_{n} a_n^{\pm} \mathcal{Z}_{(n)}$ can be expressed as a single function of the Riemann tensor. Formulating a single Lagrangian in the framework of QT gravity to derive these bounce solutions remains a topic for future investigation.

\section*{Acknowledgments}
We are very grateful to Bin-ye Dong, Fang-jing Cheng, Kai Li, Pan Li, Wen-bin Pan for helpful discussions, specially to Cong Zhang for his respondence with very helpful suggestions. This work is supported in part by the Natural Science Foundation of China (Grant Nos.~12035016,~12275275).

\bibliography{referrences}

\end{document}